\documentclass[twocolumn,reprint,aps,floatfix,dvipdfmx,longbibliography]{revtex4-2}
\usepackage{amsmath,amssymb}
\usepackage{graphicx}
\usepackage{bm}
\usepackage{braket}
\usepackage[ISO]{diffcoeff}
\usepackage{physics2}
\usepackage{xcolor}
\usepackage{dcolumn}
\usepackage{ulem}
\usepackage{mhchem}
\usepackage{siunitx}
\usepackage[
colorlinks=true, citecolor=blue, urlcolor=blue, linkcolor=blue,
setpagesize=false, bookmarks=false
]{hyperref}

\usephysicsmodule{ab}

\begin{document}

\title{
  DC electric field driven discretization of single-particle excitation spectra in a Mott insulator
}

\author{Koudai Sugimoto}
\email{sugimoto@rk.phys.keio.ac.jp}
\affiliation{
Department of Physics, Keio University, Yokohama 223-8522, Japan
}

\date{\today}

\begin{abstract}
We theoretically investigate the single-particle excitation spectra of a one-dimensional Hubbard model at half filling using an infinite matrix-product state and elucidate the discretized energy spectra emerging under the influence of a dc electric field.
In a weak electric-field regime, we observe two kinds of spectral structures in the density of states.
With increasing the electric-field strength, the discretized spectra, the period of which is proportional to the strength, become dominant, and the density of states exhibits the Wannier-Stark ladder in their spectra.
In addition, we also simulate time- and angle-resolved photoemission spectroscopy using an ultrashort terahertz pump pulse that approximates a dc electric field.
Our results represent a significant step forward in understanding the states in strongly correlated electron systems driven by a static electric field.
\end{abstract}

\maketitle

\section{Introduction}

One of the central issues in quantum physics is understanding how electrons behave under the influence of an external electromagnetic field.
Historically, phenomena such as Landau quantization in a magnetic field \cite{Landau1930ZP} and the Stark effect in an electric field \cite{Stark1913Nature} have been extensively discussed.
In 1960, Wannier proposed the so-called Wannier-Stark effect, wherein the wave function of electrons in a periodic potential undergoes localization when subjected to a tilted potential, achievable through a static electric field \cite{Wannier1960PR, Wannier1962RMP, Gluck2002PR}.
This prediction was later experimentally observed in the discretized optical spectra associated with the localization, using semiconductor superlattices \cite{Mendez1988PRL, Voisin1988PRL, Leo1992SSC}.
Come to think of it, a crystal is a representative example of where periodic potential naturally appears.
Although observing the Wannier-Stark effect in a crystal is challenging since a strong electric field is required for the short period of the potential, recent advancements in ultrashort and high-intensity terahertz light have enabled the observation of discretized optical spectra in materials \cite{Schmidt2018NC, Berghoff2021NC}.

Beyond semiconductor cases, another important question concerns the behavior of strongly correlated systems, where quantum many-body effects are intrinsic, under an electric field.
In Mott insulators, interacting electrons yield a finite energy gap.
Previous studies employing the nonequilibrium Green's function have clarified that the Wannier-Stark effect can manifest as a discretized density of states (DOS) \cite{Joura2008PRL, Tsuji2008PRB, Eckstein2011PRL, Neumayer2015PRB} and relating to it the dielectric breakdown under a strong electric field has been discussed \cite{Aron2012PRB, Eckstein2013JPCS, Lee2014PRB, Murakami2018PRB}.
Note that there are also discussions on the mechanism of dielectric breakdown by exact manners such as the Bethe ansatz and/or the density-matrix renormalization group \cite{Oka2003PRB, Oka2005PRL, Oka2010PRB, Eckstein2010PRL, Kirino2010JPSJ, Oka2012PRB}.
From the perspective of statistical physics, the similarity between the localization of interacting systems under a tilted potential and many-body localization has recently been highlighted \cite{Schulz2019PRL, Nieuwenburg2019PNAS, Scherg2019NC} and attracted considerable attention.

In this study, we focus on the single-particle excitation spectra of the half-filled one-dimensional Hubbard model, a representative model of strongly correlated systems, under the influence of a dc electric field.
Most previous studies employ dynamical mean-field theory to address electron correlation effects in the excitation spectra.
By utilizing the infinite time-evolving block decimation (iTEBD) method \cite{Vidal2007PRL, Orus2008PRB}, we can represent a time evolution of a quantum state in the thermodynamic limit in an unbiased manner.
Previously, we have investigated the Wannier-Stark ladder (WSL) appearing in the spectra of optical conductivity by using iTEBD \cite{Udono2023PRB}, which demonstrates the usefulness of this method for this purpose.

\section{Model and method}

\subsection{One-dimensional Hubbard model in a dc electric field}

The Hamiltonian of the half-filled one-dimensional Hubbard model is expressed as
\begin{align}
  \hat{H}
    &= -t_{\text{h}} \sum_{j, \sigma}
      {\hat{c}^{\dagger}_{j, \sigma} \hat{c}_{j+1, \sigma} + \text{H.c.}}
\notag \\
    &\quad + U \sum_{j}
      \pab{\hat{n}_{j,\uparrow} - \frac{1}{2}}
      \pab{\hat{n}_{j,\downarrow} - \frac{1}{2}},
\label{eq:Hamiltonian_of_1DHM}
\end{align}
where $\hat{c}_{j,\sigma}$ ($\hat{c}^\dagger_{j,\sigma}$) is the annihilation (creation) operator of an electron at site $j$ with spin $\sigma$, $t_{\text{h}}$ is the hopping integral, and $U$ is the on-site interaction.
The number operators of the electrons are defined as $\hat{n}_{j,\sigma} = \hat{c}^{\dagger}_{j,\sigma} \hat{c}_{j,\sigma}$.
In subsequent discussions, we use $U/t_{\text{h}} = 8$.

Let us consider the quantum state of this system when a dc electric field with strength $E_0$ is applied using the iTEBD method.
When the lattice constant is $a$ and the electron charge is $q$, the electric-field term is given by $H' = q \sum_{j} \phi_j \hat{n}_j$, where $\phi_j = -j a E_0$ is a scalar potential at site $j$.
Unfortunately, the iTEBD method cannot treat this form since it relies on translational symmetry.
Instead of using the scalar potential, we introduce the dc electric field by a spatially uniform vector potential $A(t)$.
To avoid excitations due to a nonadiabatic change in the electric field (see Appendix~\ref{sec:Time-dependence-of-the-single-particle-excitation-spectra}), we employ a smoothly varying vector potential given by
\begin{equation}
  A(t) = 
  \begin{cases}
    0 & (t \leq 0) \\
    -\frac{E_0}{2} \bab{t - \frac{T}{\pi} \sin(\frac{t \pi}{T})} &(0 < t < T)\\
    -E_0 \pab{t - \frac{T}{2}} & (t \geq T)\\
  \end{cases}.
\label{eq:dc-vector-potential}
\end{equation}
The electric field becomes $E(t) = E_0$ for $t \geq T$ since $E(t) = -\difcp{A(t)}{t}$, and we set $T=20$ in this study.
Note that a gauge where the vector potential becomes zero is termed the length gauge, while one in which the scalar potential becomes zero is known as the velocity gauge.
These are interconnected through a gauge transformation \cite{Graf1995PRB}.
The vector potential is incorporated into Eq.~\eqref{eq:Hamiltonian_of_1DHM} via the Peierls substitution $-t_{\mathrm{h}} \hat{c}^{\dagger}_{j,\sigma} \hat{c}_{j+1, \sigma} \to -t_{\mathrm{h}} e^{-i a q A(t)} \hat{c}^{\dagger}_{j,\sigma} \hat{c}_{j+1, \sigma}$, and the quantum state evolves under this condition.

We set the ground state of the Hamiltonian~\eqref{eq:Hamiltonian_of_1DHM} as the initial state $\ket{\psi (t=0)}$.
Defining the time-evolution operator as $\hat{U} (t) = \hat{\mathcal{T}} \exp\pab{-i \int^{t}_{0} \hat{H}(s) \dl{s}}$ with the time-ordering operator $\hat{\mathcal{T}}$, the Heisenberg representation of an operator $\hat{O}$ is given by $\hat{O} (t) = \hat{U}^\dagger (t) \hat{O} \hat{U} (t)$.
Correlation functions are determined by employing infinite-boundary conditions (IBC) with a uniform update scheme \cite{Zauner2015JPCM}.
For a detailed description of the calculation procedure using iTEBD and IBC, see Ref.~\cite{Ejima2022PRR}.
The parameters used for the numerical calculations are listed in Appendix~\ref{sec:numerical-details}.

\subsection{Wannier-Stark effect in optical spectra}

\begin{figure}
  \begin{center}
  \includegraphics[width=\columnwidth]{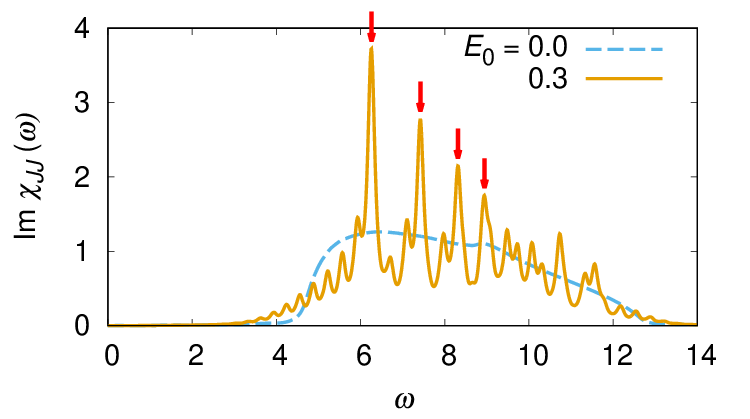}
  \end{center}
  \caption{
    The imaginary parts of the current-current correlation functions $\mathrm{Im} \, \chi_{JJ}(\omega)$ of the Hubbard model at $U=8$ for $E_0 = 0$ (blue dashed line) and 0.3 (orange solid line).
    The convergence factor is set to $\eta=0.1$.
    The red arrows indicate the significantly enhanced peaks.
  }
  \label{fig:current-correlation}
\end{figure}

It is known that a continuous energy level becomes discretized under the influence of an electric field.
Before discussing the single-particle excitation spectra, we briefly examine this effect through a dynamical current-current correlation function, which is a quantity related to optical conductivity.
Using the electric-current operator
$
\hat{J}_{A(t)}
    = - a q t_{\text{h}} \sum_{j, \sigma}
      \pab{ie^{-i a q A(t)} \hat{c}^{\dagger}_{j,\sigma} \hat{c}_{j+1, \sigma} + \text{H.c.}}
$,
the dynamical current-current correlation function can be expressed as
\begin{equation}
  \chi_{JJ}(\omega,t)
    = \frac{i}{L} \int^{\infty}_{0}
      \aab{\bab{\hat{J}_{A} (t+s), \hat{J}_{A}(t)}}
      e^{i (\omega+i\eta) s} \dl{s},
\label{eq:current-correlation}
\end{equation}
where $L$ is the number of sites and $\eta$ is a convergence factor for numerical Fourier transform.
Hereafter, we set $t_{\text{h}} = 1$, $a=1$, and $q=-1$.
If the system reaches a steady state at and after time $t$, the correlation function becomes independent of $t$.
Figure~\ref{fig:current-correlation} shows $\mathrm{Im} \, \chi_{JJ}(\omega)$.
In the absence of an electric field ($E_0 = 0$), an absorption continuum appears with a width of $8t_{\text{h}}$ the center of which is at $\omega \approx U$, which originates from the excitations from the lower-Hubbard band (LHB) to the upper-Hubbard band (UHB).
Upon applying an electric field ($E_0 = 0.3$), the continuous energy level becomes discretized, resulting in the emergence of a multiple-peak structure in the spectra.
The period of these multiple peaks is equal to $E_0$ in the limit of a strong electric field.
The origin of these discretized peaks is discussed in detail in our previous study \cite{Udono2023PRB}.

\section{Results}

\subsection{Weak dc electric field}

\begin{figure}
  \begin{center}
  \includegraphics[width=\columnwidth]{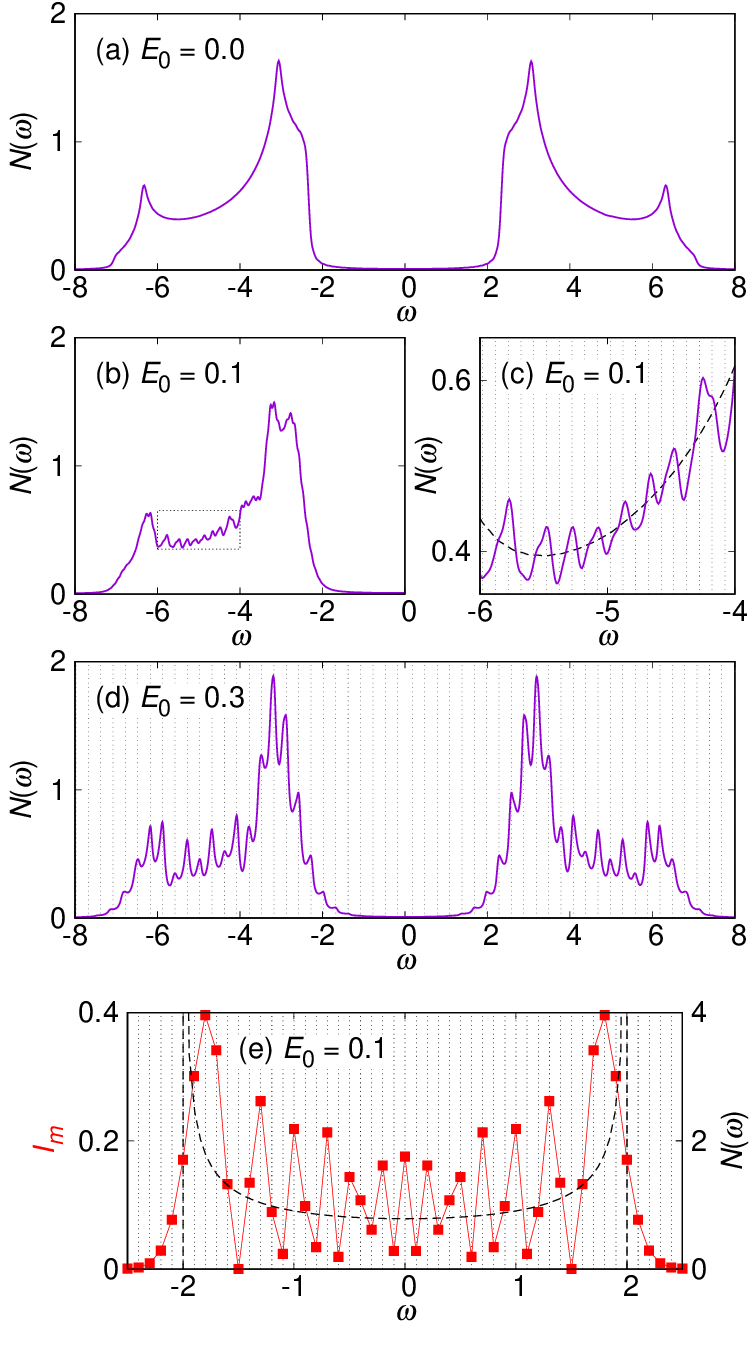}
  \end{center}
  \caption{
    The DOS $N(\omega)$ of the Hubbard model with $U=8$ under dc electric fields of (a) $E_0 = 0$, (b),(c) $E_0 = 0.1$, and (d) $E_0 = 0.3$.
    Panel (c) is a magnified view of the dotted rectangular area in Panel (b).
    The convergence factor is set to $\eta=0.05$.
    (e) Spectral weight $I_m$ at $E_0 = 0.1$ with $\omega = m E_0$ of the tight-binding model.
    The dashed black lines in Panels (c) and (e) represent the DOS in the absence of an electric field.
    The vertical dotted lines are equally distanced with a period of $E_0$.
  }
  \label{fig:spectra}
\end{figure}

We now investigate the DOS obtained from the single-particle excitation spectra.
The retarded Green's function at time $t$ is defined as \cite{Mahan2000}
\begin{multline}
  G^{\text{R}} (k,\omega, t)
    = -\frac{i}{L}
      \sum_{j,\ell}
      \int^{\infty}_{0} \dl{s}\,
      e^{-i k \cdot (r_{j} - r_{\ell}) + i(\omega+i\eta)s} 
\\
  \times
      \aab{\Bab{\hat{c}_{j,\sigma} (t+s), \hat{c}_{\ell,\sigma}^\dagger (t)}}.
\label{eq:retarded-Green-function}
\end{multline}
Using $A(k,\omega, t) = -2 \, \mathrm{Im} \, G^{\text{R}} (k,\omega, t)$, the DOS is obtained as
$
  N (\omega, t)
    = \frac{1}{2\pi} \int^{\pi}_{-\pi} A (k,\omega, t) \dl{k}.
$
For the remainder of this discussion, we omit the explicit time dependence of the DOS, as we focus on a steady state.
Figure~\ref{fig:spectra}(a) shows $N (\omega)$ in the absence of an electric field.
At half filling, an energy gap opens due to the on-site interaction $U$, giving rise to the UHB and LHB centered at $\omega \approx \pm U/2$.
Owning to the one-dimensional nature of the system, spectral peaks develop at the band edges.
Results for $A (k,\omega)$ in the Hubbard model using the iTEBD method are available in Fig.~\ref{fig:gauge-invarinat-spectra}(a) and prior studies \cite{Murakami2021PRB, Ejima2021SPP}.

Figure~\ref{fig:spectra}(b) shows $N (\omega)$ for $E_0 = 0.1$.
Notably, the band-edge peak at $\omega \approx -3$ splits into two distinct peaks.
The energy separation between these peaks significantly exceeds $E_0$, suggesting that their origin is not attributable to the formation of a WSL state.
Additionally, numerous small peaks emerge in the continuum of the Hubbard bands.
Interestingly, the spacing between these peaks deviates from $E_0$.
A magnified view of this peak structure is shown in Fig.~\ref{fig:spectra}(c), where vertical dotted lines indicate intervals of $E_0$.
The deviation of the peak separation from $E_0$ implies the emergence of an excitation structure different from the WSL.

It is worth noting that the Wannier-Stark effect always produces a series of evenly spaced peaks, separated by $E_0$, in the single-particle excitation spectra under a finite dc electric field.
The absence of such a pattern in our case can be attributed to the convergence factor $\eta$ employed in the numerical Fourier transform, which is of the same order as $E_0$.
This factor leads to the broadening of small peaks in the WSL, thereby obscuring the expected WSL structure.

Figure~\ref{fig:spectra}(d) illustrates $N (\omega)$ for $E_0 = 0.3$, where the peak separation now aligns with $E_0$, confirming the emergence of the WSL.
Notably, the peak intensities exhibit inhomogeneity.
In the optical spectra shown in Fig.~\ref{fig:current-correlation}, in addition to the equidistant peaks, we observe another set of peaks appearing above $\omega \approx 6$ (indicated by red arrows).
The periodicity of these peaks closely matches that of the pronounced peaks observed in the DOS spectra of Fig.~\ref{fig:spectra}(d).
The coexistence of two distinct peak structures in the optical conductivity has also been discussed in the previous study \cite{Chalbaud1989JPCM}.

The discretized peak structure discussed above is also observed in the spectral properties of a free fermion system. 
According to Appendix~\ref{sec:gauge-invariant-for-free-ferimons}, the DOS of a one-dimensional tight-binding model with an energy dispersion $\epsilon_k = -2 t_{\text{h}} \cos k$ [or equivalently, the case of $U=0$ in Eq.~\eqref{eq:Hamiltonian_of_1DHM}] under a finite dc electric field $E_0$ is given by
\begin{equation}
  N (\omega)
    = \sum_{m=-\infty}^{\infty} \delta \pab{\omega - i m q E_0}
    \int^{\pi}_{-\pi} J_{2m} \pab{\frac{2 \epsilon_{k}}{qE_0}} \dl{k},
\label{eq:DOS_free}
\end{equation}
where $J_n(x)$ denotes the Bessel function of the first kind of order $n$.
The delta function in Eq.~\eqref{eq:DOS_free} gives rise to the WSL structure.
Figure~\ref{fig:spectra}(e) shows the spectral weight, defined as $I_m = \int^{\pi}_{-\pi} J_{2m} \pab{\frac{2 \epsilon_{k}}{qE_0}} \dl{k}$, for $E_0=0.1$ at $\omega=mE_0$.
The spectral weight exhibits oscillations, indicating that the DOS under a dc electric field cannot be fully accounted for by the WSL picture.
In particular, spectral peaks with relatively small weight are smeared out when the influence of the convergence factor is incorporated, as demonstrated in Fig.~\ref{fig:spectra-free-fermion}(d).
Consequently, as observed in the Hubbard model [Fig.~\ref{fig:spectra}(b)], the peak spacing deviates from the expected periodicity of $E_0$.

\subsection{Strong dc electric field}
\label{sec:strong-dc-electric-field}

\begin{figure}
  \begin{center}
  \includegraphics[width=0.85\columnwidth]{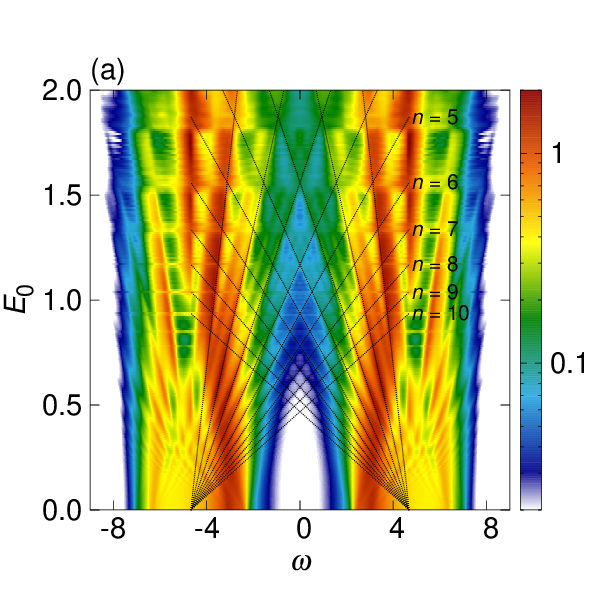}
  \\
  \includegraphics[width=0.8\columnwidth]{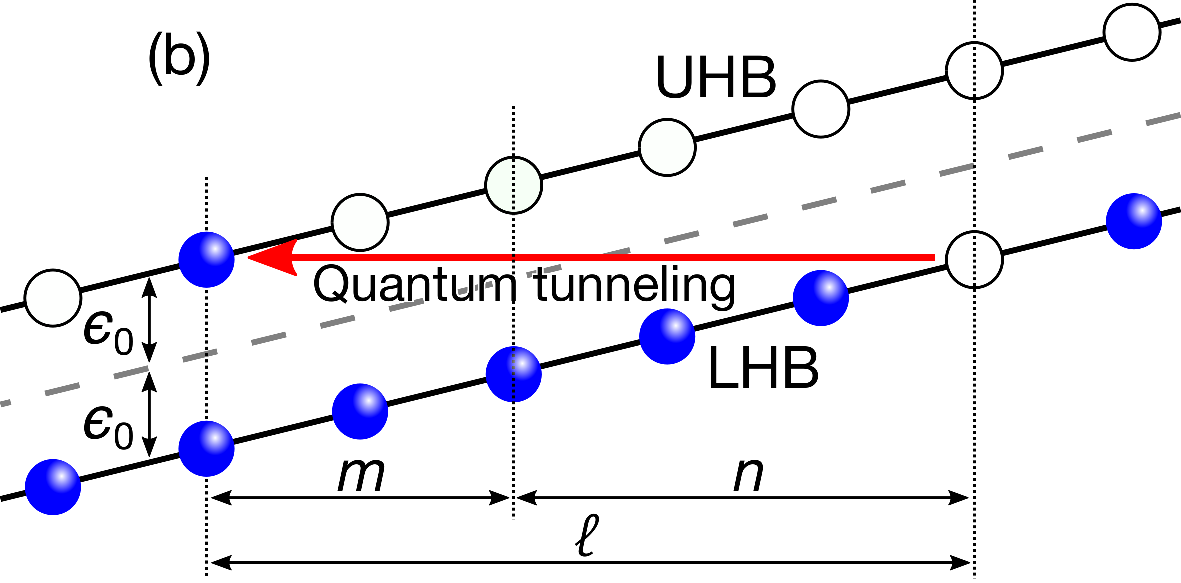}
  \end{center}
  \caption{
    (a) Contour plot of the DOS $N(\omega,t)$ in the plane of the electric-field strength $E_0$ and the frequency $\omega$ at $t=50$.
    The convergence factor is set to $\eta=0.1$.
    The black dotted lines represent the lines of $\omega = \pm \pab{\epsilon_0 - n E_0}$, where $n$ is an integer.
    (b) Schematic picture of the quantum tunneling induced by the tilted potential $V(x) = E_0 x$.
  }
  \label{fig:WSL-contour}
\end{figure}

Next, we focus on the regime with strong electric fields.
Figure~\ref{fig:WSL-contour}(a) illustrates how the DOS varies with respect to $E_0$. 
The spectra, continuous at $E_0 = 0$, become discretized with increasing $E_0$, and exhibit equidistant multiple peaks known as the WSL.
The multiple peaks are described by $\omega = \pm \pab{\epsilon_0 - n E_0}$, where $\pm \epsilon_0$ represent the energy bases of the UHB (for $+$) and the LHB (for $-$), and $n$ is an integer
\footnote{In a Mott insulator, upper- and lower-Hubbard bands exhibit approximately around $\pm U/2$ in the strong-coupling regime. We have empirically found that we can reproduce our numerical results effectively by using $\epsilon_0 = U/2 + 2 \bab{\Delta_{\text{M}} - (U/2 - 2 t_{\text{h}})} = \Delta_{\text{M}} - U/2 + 4 t_{\text{h}}$, where $\Delta_{\text{M}}$ represents the Mott gap.}.
Note that in this Mott insulator (i.e., the one-dimensional Hubbard model with $U/t_{\text{h}} = 8$), the dielectric breakdown occurs at $E_0 = 2\Delta_{\text{M}} / \xi \approx 1.668$ \cite{Oka2012PRB}, where $\Delta_{\text{M}}$ and $\xi$ represent the Mott gap and the correlation length, respectively.

Additionally, the DOS discontinuously changes at specific electric-field strengths.
These strengths can be easily estimated by considering the quantum tunneling effect.
Quantum tunneling occurs when discretized energies of the UHB and the LHB become equivalent.
This condition is expressed as $E_0 = 2\epsilon_0 / \ell$ with a positive integer $\ell$, and schematically illustrated in Fig.~\ref{fig:WSL-contour}(b).
In other words, this discontinuity appears when the WSLs from the UHB and the LHB intersect.
At the same electric-field strengths, enhancements of double occupancies are also observed, as shown in Fig.~\ref{fig:docc-and-time}.
Note that although the energy levels in Fig.~\ref{fig:WSL-contour}(b) appear unbounded due to the tilted potential, the energy range of the finite DOS does not extend to infinity.
This is because the single-particle excitation energy is measured relative to the local Fermi energy at each site.

\subsection{Simulated time- and angle-resolved photoemission spectra}

Finally, we consider time- and angle-resolved photoemission spectroscopy (TARPES) using a low-frequency pump pulse the vector potential of which is given by
$
  A(t)
    = -\frac{E_{0}}{\omega_{0}}
      e^{-(t-t_{0})^2 / 2 \sigma_{0}^2}
      \sin\bab{\omega_{0} (t-t_{0})}
$.
Here, $\omega_{0}$ is the frequency, $\sigma_{0}$ is the width, and $t_{0}$ is the central time of the pump light.
The electric field of this pump pulse reaches its maximum at $t=t_0$ with a magnitude of $E_0$.
Following the formalism proposed by Freericks \textit{et al.}, the TARPES spectrum, under the approximation that the matrix elements associated with photoemission processes are constant, is given by \cite{Freericks2009PRL, Freericks2015PS}
\begin{multline}
  I^{-} (k,\omega, t_{\text{pr}})
\\
    = \frac{1}{L}
      \sum_{j,\ell,\sigma}
      \int^{\infty}_{-\infty} \dl{t_1} \int^{\infty}_{-\infty} \dl{t_2}\,
      e^{-i \tilde{k}(t_2,t_1-t_2) \cdot (r_{j} - r_{\ell}) + i \omega (t_1 - t_2)}
\\
      \times s(t_1 - t_{\text{pr}}) s(t_2 - t_{\text{pr}})
        \aab{\hat{c}_{\ell,\sigma}^\dagger (t_2) \hat{c}_{j,\sigma} (t_1)},
\label{eq:tarpes}
\end{multline}
where $s(t-t_{\text{pr}})$ represents the envelope function of the probe pulse and
\begin{equation}
  \tilde{k} (t,s)
    = k + \frac{q}{s} \int^{t+s}_{t} A(t') \dl{t'}
\label{eq:shifted-momentum}
\end{equation}
denotes the momentum shift \cite{Boulware1966PR, Bertoncini1991PRB, Turkowski2005PRB, Tsuji2008PRB, Freericks2015PS}.
Equation~\eqref{eq:tarpes} corresponds to the imaginary part of the gauge-invariant lesser Green's function \cite{Tsuji2008PRB} incorporating the wave packets of probe pulses.
The lesser Green's function itself is a component of the retarded Green's function defined in Eq.~\eqref{eq:retarded-Green-function}.
A detailed discussion of the behavior of the gauge-invariant Green's function under a dc electric field is provided in Appendix~\ref{sec:Gauge-invariant-spectra-of-the-Hubbard-model}.
In this study, we adopt a Gaussian envelope function for the probe pulse, given by $s(t-t_{\text{pr}}) = \frac{1}{\sqrt{2\pi} \sigma_{\text{pr}}} e^{-(t-t_{\text{pr}})^2 / 2 \sigma_{\text{pr}}^2}$, where $\sigma_{\text{pr}}$ characterizes the temporal width of the probe pulse and $t_{\text{pr}}$ denotes the probe time.
Calculations of the nonequilibrium spectra using this formula in previous studies include models such as the Holstein model \cite{Sentef2013PRX}, the (extended) Hubbard model \cite{Wang2017PRB, Ejima2022PRR, Sugimoto2023PRB}, and the extended Falicov-Kimball model \cite{Ejima2022PRB}.
Hereafter, we set $\sigma_{\text{pr}} = 6$ and $t_{\text{pr}} = t_0 = 5 \sigma_0$.

\begin{figure}
  \begin{center}
  \includegraphics[width=\columnwidth]{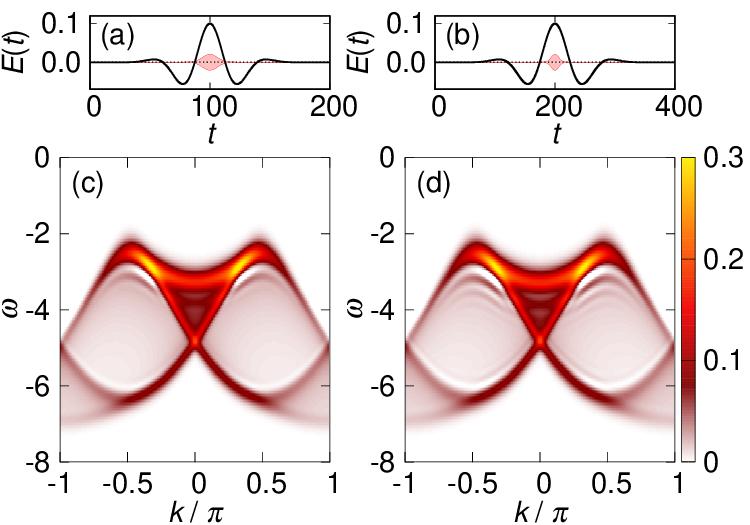}
  \end{center}
  \caption{
    (a),(b) Electric fields of pump pulse (black solid lines) and envelope curves of probe pulse with $\sigma_{\text{pr}}=6$ (red solid lines) as functions of time.
    (c),(d) TARPES spectra of the Hubbard model.
    The parameters of the pump pulse are (a),(c) $\omega_0 = 0.1$ and $\sigma_0 = 20$ and (b),(d) $\omega_0 = 0.05$ and $\sigma_0 = 40$.
    In both cases, we set $E_0 = 0.1$.
  }
  \label{fig:TARPES}
\end{figure}

Figure \ref{fig:TARPES}(c) shows $I^{-} (k,\omega, t_{\text{pr}})$ for a pump pulse with $\omega_0 = 0.1$ [see Fig.~\ref{fig:TARPES}(a)].
Notably, a new dispersion emerges just below the upper holon branch.
As observed in Fig.~\ref{fig:spectra}(b), the DOS under the dc electric field of $E_0 = 0.1$ exhibits a separation of the spectral peak at the band edge.
Since the peak position of the newly emerging dispersion almost coincides with that of the separated band-edge peak in the DOS, the calculated TARPES spectra successfully capture the characteristics of single-particle excitations.
To further refine the approximation of a dc electric field using a pump pulse, we also examine the case of $\omega_0 = 0.05$, as shown in Fig.~\ref{fig:TARPES}(d).
In this scenario, a stripe-like structure becomes apparent, suggesting a possible connection to the multiple peaks observed in the DOS.

We would like to explain how the parameters used in our current calculations correspond to actual experiments.
Materials that can be described by the one-dimensional Hubbard model include cuprates \ce{Sr2CuO3} \cite{Neudert1998PRL, Fujisawa1999PRB} and \ce{SrCuO2} \cite{Kim2004PRL}.
In \ce{SrCuO2}, for example, the on-site Coulomb interaction is estimated to be $U/t_{\text{h}} \approx 7.8$ \cite{Kim2004PRL, Benthien2007PRB}, which is close to the value used in this study.
Assuming that the hopping integral is $t_{\text{h}} \approx 0.4$~\si{eV}, the frequency of the electric field corresponds to $\omega_0 = 0.1 t_{\text{h}} \approx 10~\text{\si{THz}}$.
With a lattice constant of $a \approx 4~\mathrm{\AA}$, the estimated strength of the electric field is $E_0 = 0.1 t_{\text{h}} / a \approx 1~\text{\si{MV/cm}}$.
Since these values are practical in ultrafast terahertz experiments, the stripe structure in single-particle excitation spectra can potentially be observed using TARPES of the terahertz pump pulse.

To calculate the TARPES spectra, we have employed the two key assumptions: constant matrix elements and the momentum shift \cite{Freericks2015PS}.
As a result, the TARPES spectra obtained from Eq.~\eqref{eq:tarpes} do not strictly preserve non-negativity \cite{Tsuji2008PRB, Freericks2015PS}.
Since photoemission spectra represent probability distributions of electrons, this fact implies that the calculated TARPES spectra within this formalism are not rigorously accurate.
In our simulations, we indeed observe symptoms of negative spectral weight at larger values of $E_0$.
Ensuring positivity across a broad parameter regime would require abandoning the momentum shift approximation and incorporating momentum-dependent matrix elements, which remains an open issue.
Nonetheless, we believe that our approach remains adequate for capturing the essential qualitative features of the spectra that will be obtained experimentally.

\section{Summary}

We have investigated the single-particle excitation spectra of the one-dimensional Hubbard model under the influence of a dc electric field.
We have found that the DOS obtained from the single-particle excitation spectra exhibits two types of structures that affect the spectra of the dynamical current-current correlation function.
With increasing electric-field strength, the period of the discretized spectra in the DOS becomes proportional to this strength, corresponding to the WSL.
Within the constant-matrix approximation, we have simulated the TARPES spectra under a terahertz pump pulse, which effectively acts as a dc electric field in the electronic system.

Our analysis has demonstrated that the discretization of the energy spectra in the presence of a weak electric field cannot be solely attributed to the WSL.
The emergence of a discretized structure in the DOS is expected to be experimentally observed via photoemission spectroscopy.
We anticipate that our findings will provide valuable insights into the influence of dc electric fields on strongly correlated electron systems.

\acknowledgments
This work was supported by Japan Society for the Promotion of Science KAKENHI Grants No.~JP20H01849, No.~JP21K03439, and No.~JP23K03286, and by Japan Science and Technology Agency COI-NEXT Program Grant No.~JPMJPF2221.
The iTEBD calculations were performed using the ITensor library \cite{Fishman2022SPC}.

\appendix

\section{Numerical details}
\label{sec:numerical-details}

We perform the numerical calculations by using the infinite matrix-product state (iMPS).
The bond dimension of iMPS is set to $\chi = 1000$.
We find that the truncation error in all calculations is less than $\mathcal{O} \pab{10^{-6}}$.
To calculate the time evolution of iMPS, the second-order Suzuki-Trotter decomposition is used.
The time steps are set to $\delta t = 0.01$ for the calculations of the dynamical current correlation functions, and to $\delta t = 0.05$ for the single-particle excitation spectra.

For the numerical Fourier transform, we set the window size of the IBC and the maximum integration time to $L_{\text{w}} = 128$ (64) and $t_{\text{max}} = 100$ (50), respectively, for the convergence factor of $\eta = 0.05$ (0.1).
In addition, in Fig.~\ref{fig:TARPES}, we use $L_{\text{w}} = 64$, and set the integration ranges for $t_1$ and $t_2$ in Eq.~\eqref{eq:tarpes} to $t_{\text{pr}} \pm 3 \sigma_{\text{pr}}$.

We have carefully verified the convergence of our numerical results and confirmed that they achieve sufficient accuracy.

\begin{figure*}
  \begin{tabular}{cc}
    \begin{minipage}[t]{0.5\hsize}
      \centering
      \includegraphics[keepaspectratio, scale=0.65]{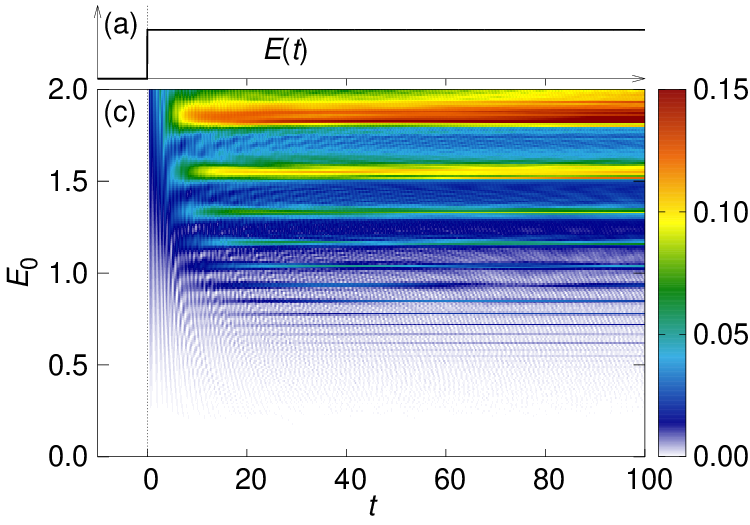}
    \end{minipage} &
    \begin{minipage}[t]{0.5\hsize}
      \centering
      \includegraphics[keepaspectratio, scale=0.65]{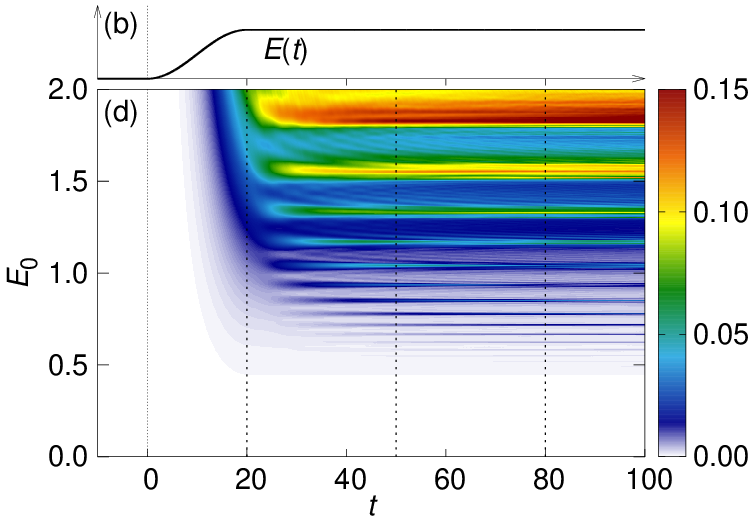}
    \end{minipage}
  \end{tabular}
  \caption{
    (a),(b) The time dependence of the electric fields given by (a) $E(t) = E_0 \theta(t)$ and (b) Eq.~\eqref{eq:dc-electric-field} with $T=20$.
    (c),(d) The time evolution of the change of double occupancies $\Delta n_{\text{d}} (t) = n_{\text{d}} (t) - n_{\text{d}} (0)$ for each case.
  }
  \label{fig:docc-and-time}
\end{figure*}

\section{Time dependence of the single-particle excitation spectra}
\label{sec:Time-dependence-of-the-single-particle-excitation-spectra}

We consider a spatially uniform electric field in the main text.
Using the gauge that the scalar potential is equal to zero, the relationship between the electric field and the vector potential is given by $E(t) = - \difcp{A(t)}{t}$.
Assuming a vector potential of $A(t) = -E_0 t \theta(t)$, the electric field becomes constant (dc electric field) and its magnitude is $E_0$ at time $t>0$.
Figure~\ref{fig:docc-and-time}(a) shows the time dependence of the electric field.
The changes of the double occupancies from its initial value for various electric-field strength in this case are illustrated in Fig.~\ref{fig:docc-and-time}(c), where the double occupancy is defined as
\begin{equation}
  n_{\text{d}} (t) = \frac{1}{L} \sum_{j}
    \aab{\hat{n}_{j, \uparrow} (t) \hat{n}_{j, \downarrow} (t)}.
\end{equation}
The double occupancy increases as the electric-field strength increases.
Certain values of $E_0$ at which the double occupancy significantly increases can be attributed to quantum tunneling, as discussed in Sec.~\ref{sec:strong-dc-electric-field} [see Fig.~\ref{fig:WSL-contour}(b)].
We find that the double occupancy does not attain a stationary value over time and continues to oscillate.
This occurs due to the abrupt application of an electric field at time $t=0$.
A sudden change of the energy of the system leads to Landau-Zener tunneling, resulting in a transition to a nonequilibrium state, and the dynamical spectra lack their stationarity.

To mitigate this effect, we employed a gradually varying electric field.
From Eq.~\eqref{eq:dc-vector-potential}, the electric field is derived as
\begin{equation}
  E(t)
  = -\diffp{A(t)}{t} =
  \begin{cases}
    0 & (t \leq 0) \\
    \frac{E_0}{2} \bab{1 - \cos(\frac{t \pi}{T})} &(0 < t < T)\\
    E_0 & (t \geq T)\\
  \end{cases}.
\label{eq:dc-electric-field}
\end{equation}
Figure~\ref{fig:docc-and-time}(b) shows the time dependence of the electric field, and corresponding changes of the double occupancy are shown in Fig.~\ref{fig:docc-and-time}(d).
The double occupancy varies smoothly with time due to the quasi-adiabatic introduction of the electric field.
Note that while the steady state is not reached at time $t=T$, the double occupancy settles to a constant value after a sufficient lapse of time.

\begin{figure*}
  \begin{center}
  \includegraphics[width=\linewidth]{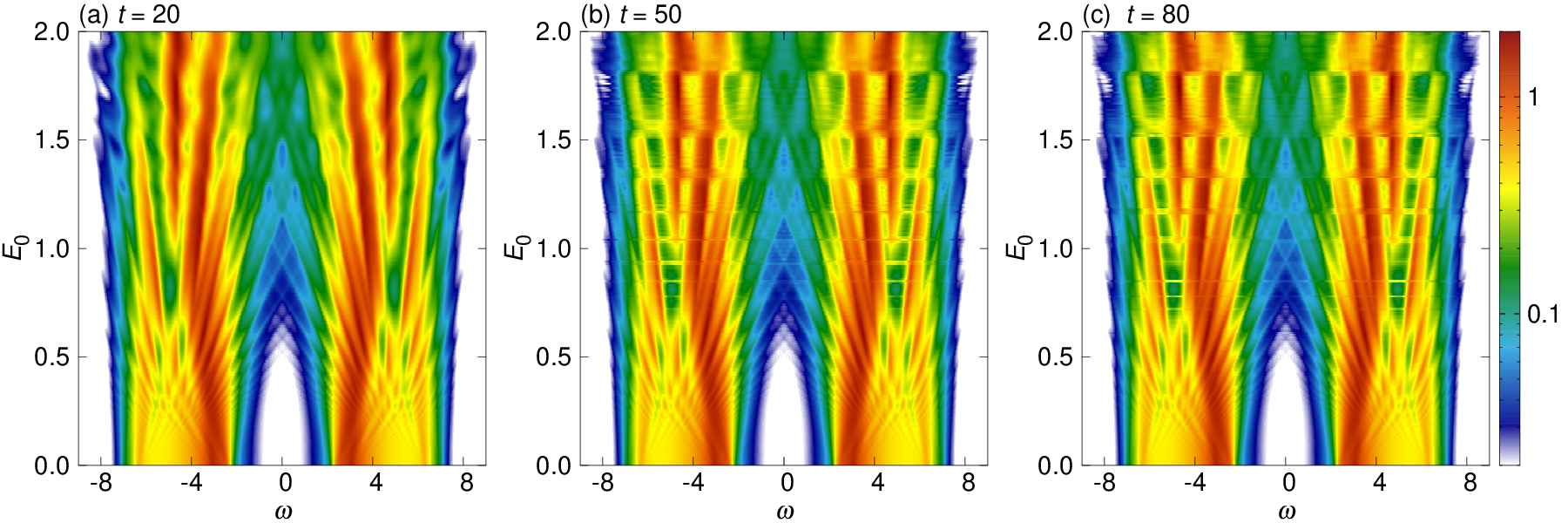}
  \end{center}
  \caption{
    Contour plots of $N(\omega, t)$ in the plane of the electric-field strength $E_0$ and the frequency $\omega$ at (a) $t=20$, (b) $t=50$, and (c) $t = 80$.
    The convergence factor is set to $\eta=0.1$.
  }
  \label{fig:dos}
\end{figure*}

Figure~\ref{fig:dos} depicts the electric-field dependence of the DOS at various times.
Upon comparing the results at $t=20$ [Fig.~\ref{fig:dos}(a)] and at $t=50$ [Fig.~\ref{fig:dos}(b)], we observe that the behaviors of the DOS are different, particularly at large values of $E_0$.
On the other hand, the result at $t=80$ [Fig.~\ref{fig:dos}(c)] is almost the same as that at $t=50$.
Thus, we conclude that the system reaches a steady state at $t=50$, and we present the results for $t=50$ in the main text.

\section{Gauge-invariant spectra of the Hubbard model}
\label{sec:Gauge-invariant-spectra-of-the-Hubbard-model}

The DOS is given by $N(\omega) = \frac{1}{2\pi} \int^{\pi}_{-\pi} A(k,\omega,t) \dl{k}$, where $A(k,\omega,t) = -2 \, \mathrm{Im} \, G^{\text{R}} (k,\omega,t)$ and $G^{\text{R}} (k,\omega,t)$ denotes the retarded Green's function defined in Eq.~\eqref{eq:retarded-Green-function}.
In general, ${G}^{\text{R}} (k,\omega,t)$ is gauge dependent.
Since the gauge field considered in the main text varies with time, $A(k,\omega,t)$ depends on time through a gauge, even in a steady state.
To address this issue, we introduce a gauge-invariant retarded Green's function $G^{\text{R}} (\tilde{k}, \omega, t)$ and a gauge-invariant spectral function $\tilde{A} (k,\omega, t) = -2 \, \mathrm{Im} \, G^{\text{R}} (\tilde{k}, \omega, t)$ \cite{Tsuji2008PRB}, where $\tilde{k}$ denotes the shifted momentum given in Eq.~\eqref{eq:shifted-momentum}.
For Eq.~\eqref{eq:dc-vector-potential} with $t \geq T$, the momentum shift is expressed as $\tilde{k} (t,s) = k + q E_0 \pab{t - \frac{T}{2} + \frac{s}{2}}$.
The DOS can be rewritten as $N(\omega) = \frac{1}{2\pi} \int^{\pi}_{-\pi} \tilde{A} (k,\omega,t) \dl{k}$.
Hereafter, we omit the explicit dependence on $t$ for the gauge-invariant quantities.

\begin{figure}
  \centering
  \includegraphics[width=\columnwidth]{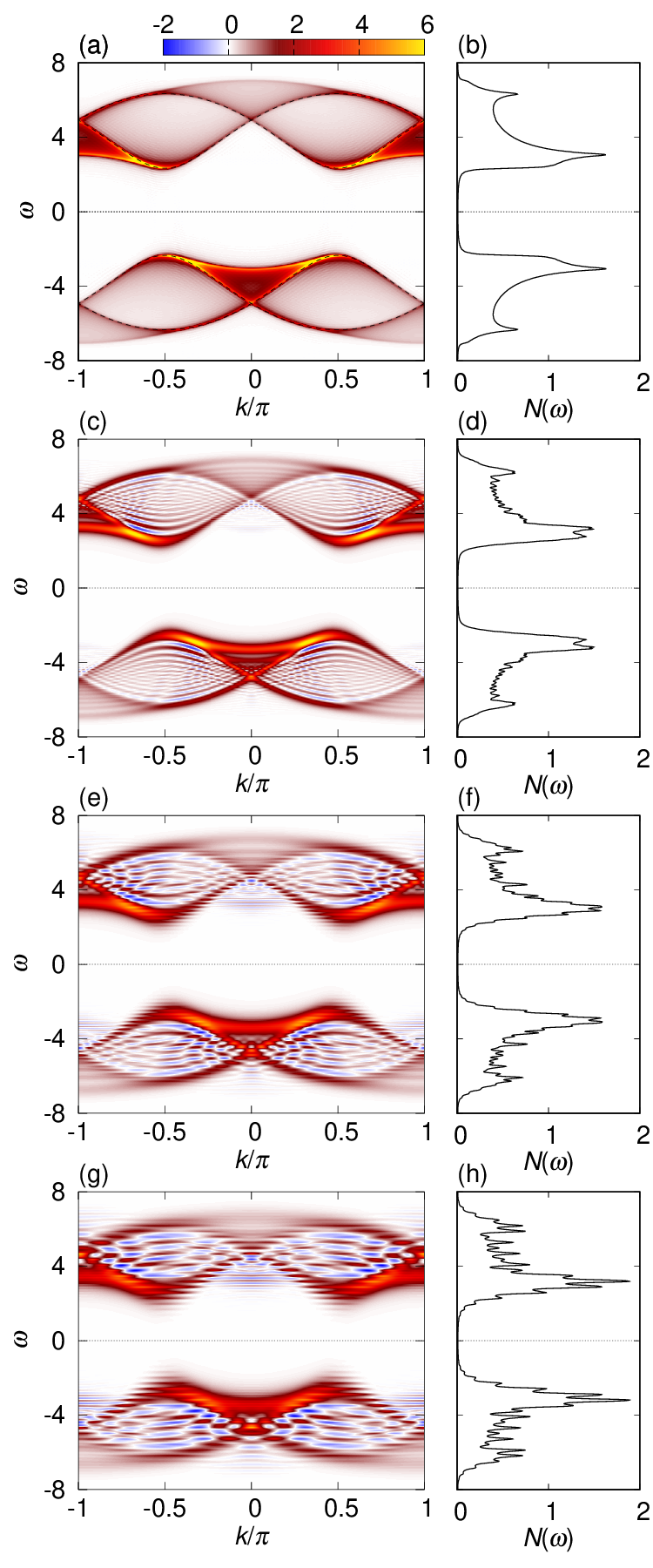}
  \caption{
    (a),(c),(e),(g) Gauge-invariant spectral functions $\tilde{A}(k,\omega)$ and (b),(d),(f),(h) the DOS $N(\omega)$ of the Hubbard model at $U=8$.
    The intensity of the electric field is set to (a),(b) $E_0 = 0$, (c),(d) $E_0 = 0.1$, (e),(f) $E_0 = 0.2$, and (g),(h) $E_0 = 0.3$.
    The convergence factor is set to $\eta=0.05$.
    The black dashed lines in Panel (a) are the holon and doublon branches obtained from the Bethe ansatz.
  }
  \label{fig:gauge-invarinat-spectra}
\end{figure}

Figure~\ref{fig:spectra}(a) shows $A (k,\omega)$ in the absence of an electric field.
The black dashed lines indicate the holon and doublon branches, which are analytically determined from the Bethe ansatz \cite{Lieb1968PRL, Essler2005}.
The corresponding DOS is presented in Fig.~\ref{fig:gauge-invarinat-spectra}(b), identical to Fig.~\ref{fig:spectra}(a).

Figure~\ref{fig:gauge-invarinat-spectra}(c) shows $\tilde{A} (k,\omega)$ for $E_0 = 0.1$.
The spectra reveal a newly emergent striped pattern within the region enclosed by the holon and doublon branches.
Notably, the periodicity of these stripes does not correspond to $E_0$, suggesting that they are not a consequence of the Wannier-Stark effect.
A similar striped pattern is observed in the spectra of a noninteracting system (see Appendix~\ref{sec:gauge-invariant-for-free-ferimons}).
Given that holons and doublons behave as free quasiparticles, it is reasonable to infer that the origin of these stripes is analogous to that in the noninteracting system.
The spectral functions $\tilde{A} (k,\omega)$ for $E_0 = 0.2$ and $0.3$ are shown in Figs.~\ref{fig:gauge-invarinat-spectra}(e) and (g), respectively.
As the field strength increases, the stripe periodicity grows larger.
Furthermore, the DOS exhibits clear spectral discretization with a periodicity of $E_0$, consistent with the Wannier-Stark effect.

In our calculations, we observe regions where $\tilde{A} (k,\omega)$ takes negative values at the tails of several newly emerged peaks.
These negative spectra arise from both the time-dependent Hamiltonian and the momentum shift.
In fact, similar negative spectra have been reported in the previous study \cite{Tsuji2008PRB}.
It is important to emphasize that, due to this issue, the conventional physical interpretation of the spectral function as a probability distribution of electrons in the energy-and-momentum space no longer holds in $\tilde{A} (k,\omega)$.
Therefore, when analyzing results based on $\tilde{A} (k,\omega)$, one must carefully assess their physical implications by comparing them with other gauge-invariant quantities.

As shown in Fig.~\ref{fig:gauge-invarinat-spectra}(h), the DOS exhibits particularly pronounced peaks at the band edges of the striped dispersions.
Moreover, the pronounced peak positions in the optical spectra align with the periodicity of the striped pattern.
Although $\tilde{A} (k,\omega)$ does not directly correspond to the angle-resolved photoemission spectra, we believe that it provides qualitative insights into the spectral properties of the system under a dc electric field.

\begin{figure}
 \centering
 \includegraphics[width=\columnwidth]{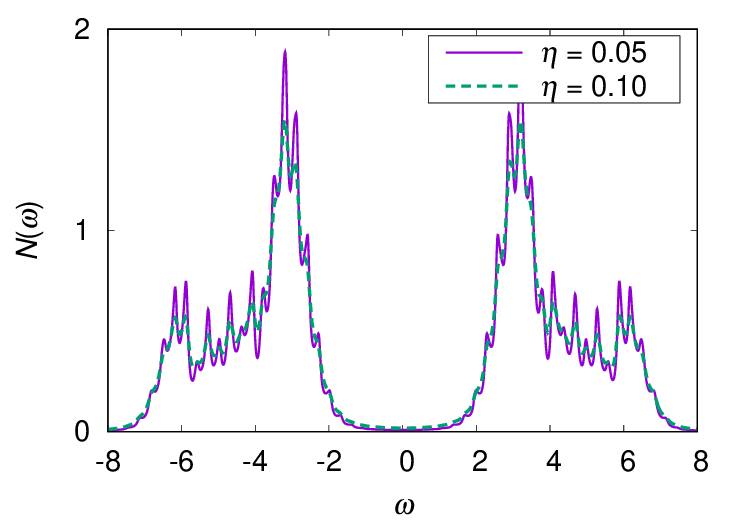}
  \caption{
    The DOS $N(\omega)$ under the dc electric field of $E_0 = 0.3$.
    The convergence factors are set to $\eta = 0.05$ (purple solid line) and $\eta = 0.1$ (green dashed line).
  }
  \label{fig:dos-E0.3}
\end{figure}

Finally, we briefly discuss the role of the convergence factor.
Figure~\ref{fig:dos-E0.3} shows the DOS for $E_0 = 0.3$ with $\eta = 0.05$ and $0.1$.
As $\eta$ increases, the spectral peaks broaden.
To clearly observe the Wannier-Stark effect, the electric-field strength $E_0$ must be significantly larger than $\eta$.
While a nonzero $\eta$ is required for the numerical Fourier transform, similar broadening effects naturally occur in real materials due to electron scattering by phonons, impurities, and other mechanisms.
In experimental settings, the effective value of $\eta$ is highly system-dependent.

\section{Gauge-invariant spectra of a noninteracting electron system}
\label{sec:gauge-invariant-for-free-ferimons}

\begin{figure}[t]
  \centering
  \includegraphics[width=\columnwidth]{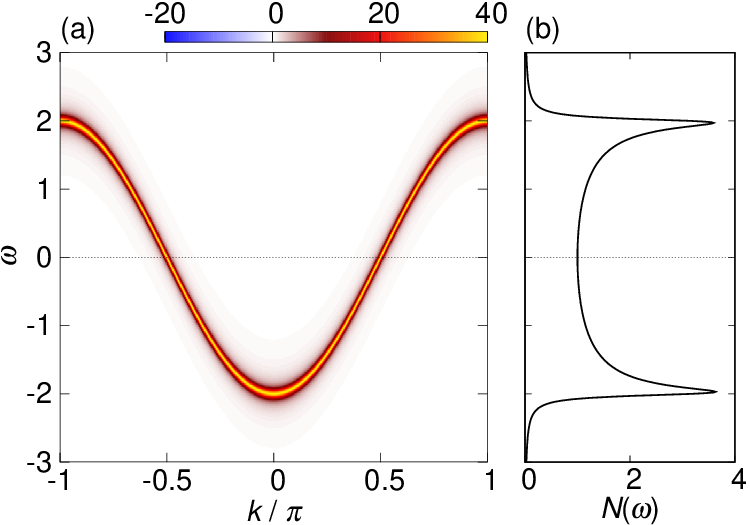}
  \includegraphics[width=\columnwidth]{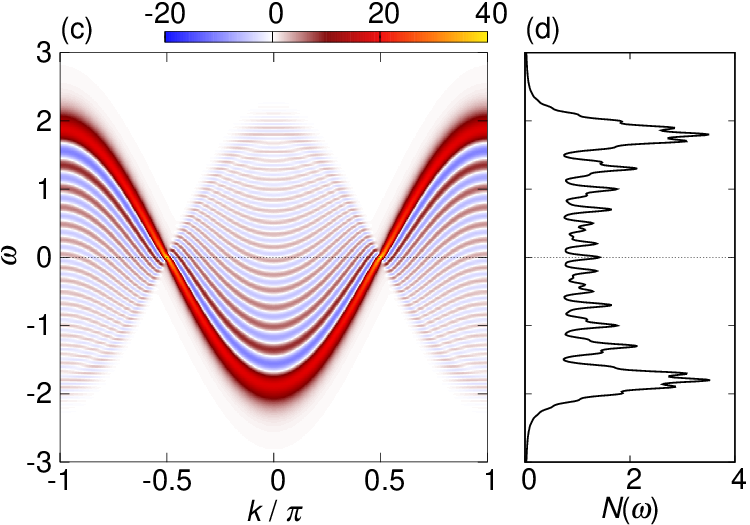}
  \caption{
    (a),(c) Gauge-invariant spectral function $\tilde{A}(k,\omega)$ and (b),(d) the DOS $N(\omega)$ of the noninteracting electron system.
    The intensity of the electric field is set to (a),(b) $E_0 = 0$ and (c),(d) $E_0 = 0.1$.
    The convergence factor in Eq.~\eqref{eq:j4Hmy6e5} is set to $\eta = 0.05$.
  }
  \label{fig:spectra-free-fermion}
\end{figure}

In the main text, we discuss the Hubbard model as a typical example of strongly correlated electron systems.
Here, we explore the spectra in a noninteracting electron system.

We consider the one-dimensional tight-binding model at half filling, the Hamiltonian of which is given by
\begin{equation}
  \hat{H}
    = \sum_{k} \epsilon_{k} \, \hat{c}^\dagger_{k, \sigma} \hat{c}_{k, \sigma},
\end{equation}
where $\epsilon_{k} = -2 t_{\text{h}} \cos k$ is the energy dispersion.
Since we focus on the case of half filling, the Fermi energy satisfies $\epsilon_{\mathrm{F}} = 0$.
Assuming the presence of the vector potential $A(t)$, the time-dependent Hamiltonian in the velocity gauge is expressed as $\hat{H} (t) = \sum_{k} \epsilon_{k - q A(t)} \, \hat{c}^\dagger_{k, \sigma} \hat{c}_{k, \sigma}$. 
By solving the Heisenberg equations of motion for the annihilation and creation operators, namely $i \diff{\hat{c}^{(\dagger)}_{k, \sigma} (t)}{t} = \bab{\hat{c}^{(\dagger)}_{k, \sigma} (t), \hat{H} (t)}$, we obtain
\begin{align}
  \hat{c}_{k, \sigma} (t)
    &= \exp \pab{-i\int^t_{-\infty} \epsilon_{k - q A(t')} \dl{t'}} \hat{c}_{\bm{k}, \sigma}
\\
  \hat{c}^\dagger_{k, \sigma} (t)
    &= \exp \pab{i\int^t_{-\infty} \epsilon_{k - q A(t')} \dl{t'}} \hat{c}^\dagger_{\bm{k}, \sigma}.
\end{align}
From the above considerations, the retarded Green's function under the influence of the vector potential is written as \cite{Turkowski2005PRB}
\begin{align}
  &G^{\text{R}} (k,t_1,t_2)
\notag \\
    &= -i \theta(t_1 - t_2) \aab{\hat{c}_{k,\sigma} (t_1) \hat{c}^{\dagger}_{k,\sigma} (t_2) + \hat{c}^{\dagger}_{k,\sigma} (t_2) \hat{c}_{k,\sigma} (t_1)}
\notag \\
    &= -i \theta(t_1-t_2) \exp \pab{-i \int^{t_1}_{t_2}
    \epsilon_{k - q A(t)} \dl{t}}.
\label{eq:DBNP4jex}
\end{align}
To render the Green's function gauge invariant, we replace $k \to \tilde{k} (t_1,t_2) = k + \frac{q}{t_1 -t_2} \int^{t_1}_{t_2} A(t) \dl{t}$, as in Eq.~\eqref{eq:shifted-momentum}.
Equation~\eqref{eq:DBNP4jex} becomes
\begin{equation}
  \tilde{G}^{\text{R}} (k,t_1, t_2)
    = i \theta(t_1-t_2)
      \exp \pab{-i \int^{t_1}_{t_2} \epsilon_{\tilde{k}(t_1,t_2) - q A(t)} \dl{t}}.
\label{eq:E8j2kLLD}
\end{equation}

Now we focus on a dc electric field.
When the magnitude of the electric field is $E_0$, the vector potential is given by $A(t) = -E_0 t \theta(t)$.
Hereafter, we consider the case where $t>0$.
Since the momentum shift required for the gauge invariance is given by
\begin{equation}
  \tilde{k} (t_1, t_2) - q A(t)
    = k + q E_0 \pab{t - \frac{t_1 + t_2}{2}},
\end{equation}
the integration appearing in the exponent in Eq.~\eqref{eq:E8j2kLLD} becomes
\begin{equation}
  \int^{t_1}_{t_2} \epsilon_{\tilde{k} (t_1, t_2) - q A(t)} \dl{t}
    = \frac{2\epsilon_{k}}{qE_0} \sin \pab{\frac{q E_0}{2} \pab{t_1 - t_2}}.
\end{equation}
Using this, the gauge-invariant Green's function in Eq.~\eqref{eq:E8j2kLLD} reads
\begin{equation}
  \tilde{G}^{\text{R}} (k,t_1, t_2)
    = -i
    \exp \pab{-i \frac{2 \epsilon_{k}}{qE_0} \sin \pab{\frac{q E_0}{2} \pab{t_1 - t_2}}}.
\label{eq:U2R29HPE}
\end{equation}
Since the Green's function is a function of $t_1-t_2$, we write the Green's function as $\tilde{G}^{\text{R}} (k,t = t_1-t_2)$ for simplicity.
The Fourier transform of the Green's function is
\begin{align}
  \tilde{G}^{\text{R}} (k, \omega)
    & = \int^{\infty}_{-\infty} \tilde{G}^{\text{R}} (k, t) e^{i \omega t} \dl{t}
\notag \\
    &= -i \int^{\infty}_{0}
      \exp \pab{i \bab{\omega t - \frac{2 \epsilon_{k}}{qE_0} \sin \pab{\frac{q E_0}{2} t}}} \dl{t},
\end{align}
and applying the formula on the Bessel function, $e^{i x \sin \theta} = \sum_{n=-\infty}^{\infty} J_n(x) e^{in\theta}$, we obtain
\begin{equation}
  \tilde{G}^{\text{R}} (k, \omega)
    = \sum_{n=-\infty}^{\infty}
    J_n \pab{\frac{2 \epsilon_{k}}{qE_0}}
    \frac{1}{\omega + i\eta - in\frac{q E_0}{2}}.
\label{eq:j4Hmy6e5}
\end{equation}
Therefore, the gauge-invariant Green's function at $\eta \to 0$ is given by
\begin{align}
  \tilde{A} (k, \omega)
    &= -2 \, \mathrm{Im} \, \tilde{G}^{\text{R}} (k, \omega)
\notag \\
    &= 2 \pi \sum_{n=-\infty}^{\infty}
    J_n \pab{\frac{2 \epsilon_{k}}{qE_0}}
    \delta \pab{\omega - in\frac{q E_0}{2}}.
\end{align}
Due to the presence of the delta function, the electric field discretizes the energy levels of the gauge-invariant spectral function in increments of $qE_0/2$.
On the other hand, since the energy dispersion $\epsilon_k$ is an even function and the Bessel function is an odd function for odd $n$, the DOS, $N(\omega) = \frac{1}{2\pi} \int^{\pi}_{-\pi} A(k,\omega)\dl{k}$, is discretized in units of $qE_0$.
As a result, we finally obtain Eq.~\eqref{eq:DOS_free}.
This phenomenon corresponds to the Wannier-Stark effect.

In addition to the discretized structure, which is equally spaced and proportional to the electric field in the energy spectra, the spectral function exhibits the oscillatory behavior attributable to the Bessel function.
Since the Bessel function can take negative values, the gauge-invariant spectral function inherently includes negative regions.
This fact indicates that the gauge-invariant spectral function does not represent a probability distribution of electrons in the energy-and-momentum space.

Figure~\ref{fig:spectra-free-fermion} displays $\tilde{A}(k,\omega)$ and $N(\omega)$ for $E_0 = 0$ and $0.1$.
As shown in Fig.~\ref{fig:spectra-free-fermion}(a), prominent peaks appear along the energy dispersion $\omega = \epsilon_k = -2 \cos k$.
Reflecting its one-dimensional nature, the DOS presented in Fig.~\ref{fig:spectra-free-fermion}(b) exhibits peaks at the band edges.

In the gauge-invariant spectral function for an electric field of $E_0=0.1$, shown in Fig.~\ref{fig:spectra-free-fermion}(c), a stripe structure emerges alongside the original band.
At the same time, regions where the spectra become negative appear.
Note that since the DOS is obtained by integrating over momentum, the problem arising from the gauge field does not occur.
The DOS shown in Fig.~\ref{fig:spectra-free-fermion}(d) exhibits multiple peaks, the number of which is consistent with the number of bands forming the stripe structure.
This observation suggests that the stripe structure---excluding the regions of negative spectral weight---should be experimentally observable in angle-resolved photoemission spectroscopy.

In the case of the Hubbard model, we observe the emergence of the stripe structure in the regions surrounded by the holon and doublon bands.
While explaining the exact origin of the striped spectra may be challenging, considering that holons and doublons can be treated as free quasiparticles, we believe that the origin of the stripe structure may be similar to that observed in the noninteracting electron system.

\bibliography{paper.bbl}

\end{document}